\documentclass[12pt]{iopart}
\usepackage{iopams}
\usepackage{epsfig}

\begin{document}

\tolerance = 10000

\title[Trapped interacting two-component bosons]
{Trapped interacting two-component bosons in one dimension}

\author{Shi-Jian Gu$^1$, You-Quan Li$^{1,2}$ and Zu-Jian Ying$^1$}

\address{$^1$Zhejiang Institute of Modern Physics,
 Zhejiang University, Hangzhou 310027, P.R. China\\
$^2$Institute for Physics, Augsburg University,
D-86135 Augsburg, Germany}

\date{Received: June 2001}

\begin{abstract}
In this paper we solve one dimensional trapped SU(2) bosons with
repulsive $\delta$-function interaction by means of Bethe-ansatz method.
The features of ground state and low-lying excited states are
studied by numerical and analytic methods. We show  that the
ground state is an isospin ``ferromagnetic" state which differs
from spin-1/2 fermions system. There exist three quasi-particles
in the excitation spectra, and both holon-antiholon and holon-isospinon 
excitations are gapless for large systems.
The thermodynamics equilibrium of the system at finite temperature is
studied by thermodynamic Bethe ansatz. The thermodynamic
quantities, such as specific heat etc. are obtained for the case
of strong coupling limit.
\end{abstract}


\maketitle

\section{Introduction}

In the past few years, 
Bose Einstein condensation (BEC) has witnessed a sequence
of exhilarating experimental achievements, and
much attention has been paid to the study of Bose systems.
The possibility of BEC in one-dimension has been
discussed for the non-interacting Bose
gas \cite{Bagnato,Ketterle}. The role of dimensionality
has been examined for the ideal Bose gas \cite{Druten}.
It is known the interaction between
bosons plays an essential role in one dimension due to
the strong constraint in phase space \cite{LiebL}.
The Luttinger liquid properties \cite{Haldane} of trapped, interacting
quasi-one-dimensional Bose gas was discussed \cite{Monien}.
It was shown under suitable experimental conditions that the system
can be described as a Luttinger liquid and the correlation
function of the bosons decays algebraically which prevents
Bose-Einstein condensation.
Recently, a two component Bose gas was produced in magnetically
trapped $^{87}Rb$ by rotating the two hyperfine states into each
other with the help of slightly detuned Rabi oscillation
field \cite{Williams1,Williams2}. It was noticed \cite{Ho} that the
properties of the Bose system can be different from the traditional
scalar Bose system once it acquires internal degree of freedom.
The two-component Bose system on a circle (periodic boundary conditions)
has been studied by means of Bethe-ansatz method \cite{LiGYE}.
It was shown that the ground state is an isospin-ferromagnetic state
which differs from the one-dimensional spin-1/2 Fermi system
whose ground state is SU(2) singlet.

In this paper, we solve trapped two-component bosons with $\delta$-function
interaction in one dimension by Bethe-ansatz method.
On the basis of Bethe-ansatz equation, we discuss the ground state,
low-lying excited states and the thermodynamics
of the system at finite temperature,
where some thermal quantities at low temperature are obtained explicitly. 
Our paper is organized as follows: In the next section we
introduce the model and derive a set of non-linear equations
for charge rapidity and isospin rapidity.
In Sec. \ref{sec:ground}, we explicitly show that the ground state
is an isospin ``ferromagnetic" state and manifest how the quantum
numbers in Bethe-ansatz equation should be taken for the ground
state. In Sec. \ref{sec:low}, We study the low-lying excited states extensively
by analyzing the possible variations in the sequence of quantum numbers.
Numerical results of energy-``momentum'' spectra for each excitation are
given. In Sec. \ref{sec:thermo}
we discuss the general thermodynamics of the system by means of
thermodynamic Bethe ansatz \cite{Yang}.
In Sec. \ref{sec:spec} we derive the free energy
and specific heat in the case of strong coupling limit.
In the last section a brief summary is given.

\section{The model and its Bethe-ansatz solution}\label{sec:model}

The Hamiltonian for a two-component bosons trapped  in
a potential well of infinite depth reads
\begin{equation}
H=-\sum_{i=1}^N \frac{\partial^2}{\partial x_i^2}
  +\sum_{i=1}^N V(x_i)+2c\sum_{i>j=1}^N\delta(x_i-x_j)
\label{eq:Hamiltonian}
\end{equation}
where
\begin{equation}
V(x_i)=\left\{
\begin{array}{cc}
0\;\;\;\;\;\;|x|\leq L/2\\
\infty\;\;\;\;\;|x|>L/2
\end{array}
\right.
\nonumber
\end{equation}
The Hamiltonian is not invariant under
translation due to the presence of trapping potential,
and the total momentum of the system
is not conserved. However, the system is still
invariant under the action of the permutation group
$S_N$ which makes it possible to employ the
coordinate Bethe-ansatz approach.

In the domain with $x_i\neq x_j(i\neq j)$ and inside the potential well,
the Hamiltonian reduces to the one for free bosons and it's eigenfunctions
are therefore just the superpositions of
plane waves. When two particles collide with each other at the
same point, a scattering process happens. The Bethe-ansatz embodies that
this process is purely elastic and the occurrence during the process is merely
that the particles exchange their momenta. So for a given momentum
$k=(k_1, k_2,\dots,k_N)$, the scattering momenta include
all permutations of the components of $k$. Moreover,
since the total momentum is not conserved in the present
model, we will have more nondiffractive scattering momenta,
e.g., for $N=2$, we have $(k_1, k_2)$ $\rightarrow$ $(-k_1, k_2)$,
$(k_1, -k_2)$, $(-k_1, -k_2)$, $(k_2, k_1)$, $(-k_2, k_1)$,
$(k_2, -k_1)$, or $(-k_2, -k_1)$.
All the eight states correspond to the same
energy $k_1^2+k_2^2$. Hence, besides all possible permutations
of the components of $k$, the scattering must include all
possibilities of sign changes in the components of $k$.

Now we consider the case of $N$ bosons.
Considering the fact that the total momentum is not conserved, we adopt
the following Bethe-ansatz  form \cite{Li}
\begin{equation}
\psi_a (x)=\sum_{P\in {\cal W_B}}A_a(P,Q)e^{i(P k|Qx)}
\; x \in {\cal C}(Q)
\label{eq:BAWF}
\end{equation}
where
$a=(a_1, a_2,\ldots,a_N)$, $a_j$ denotes the isospin label of the $j$th
particle;
$Pk $ stands for the image of a given
$ k:= ( k_1 , k_2, \cdots , k_N ) $
by a mapping $P\in {\cal W}_{B} $
and the coefficients $ A(P,Q)$
are functionals on ${\cal W}_{B} \otimes {\cal W}_{A}$.
Here, ${\cal W}_B$ and ${\cal W}_A$ stand for the Weyl group of $B_N$ and
$A_{N-1}$ Lie algebras respectively. The later is isomorphic to the
permutation group $S_N$ and the former consists of $S_N$ and all
possible sign changes.
We emphasize that the sum runs over the Weyl group of
Lie algebra $B_N $ but the wave function is defined on various
Weyl chambers ${\cal C}(Q)$
corresponding to the Weyl group of $A_{N-1}$ Lie algebra.
This is different from the situation of periodic boundary conditions.

For a Bose system, the wave function is supposed to be
symmetric under any permutation of  both  coordinates and
isospin indices, i.e.
\begin{equation}
(\sigma^j\psi)_a (x)=\psi_a(x)
\label{eq:permutation}
\end{equation}
Here $(\sigma^j \psi)_a$ is well defined by
$\psi_{\sigma^j a}(\sigma^j x)$,
therefore both sides of (\ref{eq:permutation}) can be written out by
using (\ref{eq:BAWF}).
Furthermore, using the evident identity
$ (Pk|\sigma^i Qx) = (\sigma^i Pk|Qx)$
and the rearrangement theorem of group theory, we obtain the
following consequence from (\ref{eq:permutation})
\begin{equation}
A_a(P,\sigma^i Q)=A_{\sigma^j a}(\sigma^i P,Q).
\label{eq:symmetry}
\end{equation}

The $\delta$-function term in the Hamiltonian (\ref{eq:Hamiltonian})
contributes a boundary conditions
at hyperplane $ {\sf l \! P }_{\alpha} $ ($\alpha$ is a root of Lie
algebra $A_{N-1}$), namely a discontinuity
of the derivatives of the
wave function along the normal of a Weyl hyperplane:
\begin{equation}
\lim_{\epsilon\rightarrow 0^+}
 [\alpha\cdot\nabla\psi_a(x_{(\alpha)}+\epsilon\alpha)
  -\alpha\cdot\nabla\psi_a(x_{(\alpha)}-\epsilon\alpha)]
  =2c\psi_a(x_{(\alpha)})
\label{eq:discontinuity}
\end{equation}
where $x_{(\alpha)} \in {\sf l \! P }_{\alpha} $
and $ \nabla := \sum^{N}_{i=1} e_i (\partial / \partial x_i )$.
Substituting (\ref{eq:BAWF}) into (\ref{eq:discontinuity}), we find that
\begin{eqnarray}
i[(Pk)_j-(Pk)_{j+1}]
[A_a(P,\sigma_iQ)-A_a(\sigma_iP,\sigma_iQ)
-A_a(P,Q)+A_a(\sigma_iP,Q)] \nonumber\\
 = c[A_a(P,Q)+A_a(\sigma_iP,Q)+A_a(P,Q)+A_a(\sigma_i P,Q)]
\label{eq:coeff}
\end{eqnarray}
By making use of the relation (\ref{eq:symmetry}) and the continuity relation
of wave function across $(Qx)_i=(Qx)_{i+1}$,
we can obtain the following relations
\begin{eqnarray}
A_a(\sigma_iP,Q)=\check{S}_{a,a'}^i(P k)A_{a'}(P,Q)
\label{eq:Arelation} \\
\check{S}_{a,a'}^i(P k)= -
\frac{c \delta_{a, a'}+i[(P k)_i-(P k)_{i+1}\,] {\cal P}_{a,a'}}
     {c-i[(P k)_i-(P k)_{i+1}] \,}
\label{eq:Smatrix}
\end{eqnarray}
where ${\cal P}_{a, a'}$ stands for the matrix elements of
the spinor representation of the permutation group.
The relation (\ref{eq:permutation}) provides  for the coefficients
$A$ a relation between different Weyl chambers.
Eq.(\ref{eq:Arelation}) provides a connection between those coefficients which
are related via any element of Weyl group ${ \cal W}_{B} $ in the same
Weyl chamber.

The basic elements of the Weyl group ${ \cal W}_{B} $ obey
$\sigma^i\sigma^i = 1$  and
$\sigma^{i} \sigma^{i+1} \sigma^{i}=
\sigma^{i+1} \sigma^{i} \sigma^{i+1}$
as identities. These identities imply that
$
A_a(\sigma^i\sigma^i P, Q) = A_a(P, Q)
$ and
$
 A_a(\sigma^{i}\sigma^{i+1}\sigma^{i}P, Q)=
 A_a(\sigma^{i+1}\sigma^{i}\sigma^{i+1}P, Q)
$.
Using (\ref{eq:Arelation}) repeatedly, one can obtain the following relations:
\begin{flushleft}
$\check{S}^i (\sigma^i P k)\check{S}^i (\sigma k)=I$
\nonumber \\
$\check{S}^i (\sigma^{i+1}\sigma^i Pk)
\check{S}^{i+1}(\sigma^i Pk)
\check{S}^i (Pk) =
\check{S}^{i+1}(\sigma^i \sigma^{i+1} Pk)
\check{S}^i (\sigma^{i+1} Pk)
\check{S}^{i+1}(Pk)$
\end{flushleft}
\begin{equation}
\label{eq:ci}
\end{equation}
where we have adopted the conventions $ \check{S}=mat(S_{ab})$,
$\check{S}^i = \check{S}\otimes I, \,
\check{S}^{i+1}= I\otimes\check{S}\,$
($I$ is a $2\times 2$ unit matrix).
These relations are consistency conditions for the S-matrix.
The second relation is called  the Yang-Baxter equation.
The concrete S-matrix in (\ref{eq:Smatrix}) fulfills these relations.

Due to the infinite depth of the potential well, the wave function
must vanish at the ends of the well,
\begin{equation}
\psi_a \bigl((Qx)_1=-L/2\bigr)=0,\;\;
\psi_a\bigl( (Qx)_N=L/2\bigr)=0
\label{eq:boundary}
\end{equation}
These boundary conditions give rise to
\begin{eqnarray}
A_a(\tau^1 P,Q)=-e^{-i(Pk)_1L}A_a(P,Q)
 \label{eq:left} \\
A_a(\gamma P,Q)=-e^ {i(Pk)_NL}A_a(P,Q)
\label{eq:right}
\end{eqnarray}
where $\gamma=\sigma^{N-1}\ldots\sigma^1\tau^1\sigma^1\ldots\sigma^{N-1}$.
The relation (\ref{eq:left}) together with eq.(\ref{eq:Arelation})
determine the amplitudes $A$'s up to an overall factor since
$\sigma^1,\ldots,\sigma^{N-1}$ and $\tau^1$ generate the whole ${\cal W}_B$.
There are two further consistency conditions. One comes from the
identity ($\tau^1\sigma^1\tau^1\sigma^1=\sigma^1\tau^1\sigma^1\tau^1$)
of the Weyl group ${\cal W}_B$ leading to a reflection related Yang-Baxter
equation 
which is fulfilled automatically due to the fact that the reflection matrix
is just the unit matrix multiplied by a scalar function.
The other one comes from eq.(\ref{eq:right}) which leads to an eigenvalue
equation for the products of the S-matrices. This eigenequation can be
diagonalized by means of standard quantum inverse scattering method\cite{LDFaddeev84}.
One finally obtains the following Bethe-ansatz equations
\begin{eqnarray}
e^{i2k_jL}=\prod_{l\neq j}^N
\frac{k_j-k_l+ic}{k_j-k_l-ic}
 \frac{k_j+k_l+ic}{k_j+k_l-ic}
  \prod_{\mu=1}^{M}
   \frac{k_j-\lambda_\mu-ic/2}{k_j-\lambda_\mu+ic/2}
    \frac{k_j+\lambda_\mu-ic/2}{k_j+\lambda_\mu+ic/2}
     \nonumber\\
1=\prod_{l=1}^{N}
\frac{\lambda_\gamma-k_l-ic/2}{\lambda_\gamma-k_l+ic/2}
 \frac{\lambda_\gamma+k_l-ic/2}{\lambda_\gamma+k_l+ic/2}
  \prod_{\mu\neq\gamma}^M
   \frac{\lambda_\gamma-\lambda_\mu+ic}{\lambda_\gamma-\lambda_\mu-ic}
    \frac{\lambda_\gamma+\lambda_\mu+ic}{\lambda_\gamma+\lambda_\mu-ic}
\label{eq:BAE}
\end{eqnarray}
where $M$ denotes the total number of down isospins and $\lambda$ denote
isospin rapidities which arise from the diagonalization conditions of
quantum inverse scattering method.
Taking the logarithm of eqs. (\ref{eq:BAE}) gives rise to the
secular equations:
\begin{flushleft}
$\displaystyle
\pi I_j =k_j L+
  \sum_{l\neq j}^N \Bigl[\tan^{-1}\Bigl(\frac{k_j-k_l}{c}\Bigr)
   +\tan^{-1}\Bigl(\frac{k_j+k_l}{c}\Bigr)\Bigr]
   -\sum_{\mu=1}^M\Bigl[\tan^{-1}\Bigl(\frac{k_j-\lambda_\mu}{c/2}\Bigr)
    +\tan^{-1}\Bigl(\frac{k_j+\lambda_\mu}{c/2}\Bigr)\Bigr]$  \\[3mm]
$\displaystyle
\pi J_\gamma=
  \sum_{l=1}^N\Bigl[\tan^{-1}\Bigl(\frac{\lambda_\gamma-k_l}{c/2}\Bigr)
  +\tan^{-1}\Bigl(\frac{\lambda_\gamma+k_l}{c/2}\Bigr)\Bigr]
  -\sum_{\mu \neq \gamma}^M \Bigl[\tan^{-1}
   \Bigl(\frac{\lambda_\gamma-\lambda_\mu}{c}\Bigr)
   +\tan^{-1}\Bigl(\frac{\lambda_\gamma+\lambda_\mu}{c}\Bigr)\Bigr]$
\end{flushleft}
\begin{equation}\label{eq:secular}\end{equation}
where $I_j$ is the quantum number for the
charge rapidity $k_j$ and $J_\gamma$ for the isospin rapidity.
Concerning the property of the logarithm
function, $I_j$ and $J_\gamma$ take integer values
regardless of either the total number of particles
or that of isospin down.
Once the roots are solved
from eqs. (\ref{eq:secular}) for a given
set of quantum numbers $\{I_j, J_\gamma\}$, the energy can
be calculated by $E=\sum_{j=1}^N k_j^2$.

\section{The ground state}\label{sec:ground}

For any set of quantum numbers $\{I_j, J_\gamma\}$ with the solution
$\{ k_j, \lambda_\gamma\}$,
the replacement of either $I_j\rightarrow -I_j$, $k_j\rightarrow -k_j$
or $J_\gamma\rightarrow -J_\gamma$, $\lambda_\gamma\rightarrow -\lambda_\gamma$
keeps the secular equation (\ref{eq:secular}) invariant.
So it makes no change to the energy,
then we only need to consider the case of positive integer \cite{Gaudin}
for $Is$ and $Js$.
In the weak coupling limit $c\rightarrow 0$,
$\tan^{-1}(x/c)\rightarrow \pi{\rm sgn}(x)/2$,
eqs. (\ref{eq:secular}) become
\begin{eqnarray}
\pi I_j=k_j L
  +\frac{\pi}{2}\sum_{l\neq j}^N
   [{\rm sgn}(k_j-k_l)+{\rm sgn}(k_j+k_l)]
   -\frac{\pi}{2}\sum_{\mu=1}^M[{\rm sgn}
     (k_j-\lambda_\mu)+{\rm sgn}(k_j+\lambda_\mu)]
      \nonumber \\
2J_\gamma=\sum_{l=1}^N
  [{\rm sgn}(\lambda_\gamma-k_l)+{\rm sgn}
   (\lambda_\gamma+k_l)]
   -\sum_{\mu\neq\gamma}^M
   [{\rm sgn}(\lambda_\gamma-\lambda_\mu)
   +{\rm sgn}(\lambda_\gamma+\lambda_\mu)]
\label{eq:secular-weak}
\end{eqnarray}
We may choose both the subscript of the $J_j$ and $J_\gamma$ in an
increasing order so that the second equation
of eqs. (\ref{eq:secular-weak}) becomes
\begin{equation}
\sum_{l=1}^N{\rm sgn}(\lambda_\gamma-k_l)
  =2J_\gamma+2\gamma-N-2
\label{GMM1}
\end{equation}
which also gives rise to
\begin{equation}
\sum_{l=1}^N[{\rm sgn}(\lambda_{\gamma+1}-k_l)
-{\rm sgn}(\lambda_\gamma-k_l)]
=2(J_{\gamma+1}-J_\gamma)+2
\label{GMM}
\end{equation}
Thus, for $J_{\gamma+1}-J_\gamma=m$, there must exist exactly $m+1$
momenta of $k_l$ satisfying $\lambda_\gamma<k_l<\lambda_{\gamma+1}$.

\begin{figure}
\epsfclipoff
\fboxsep=88mm
\setlength{\unitlength}{1mm}
\begin{picture}(96,42)(0,0)
\linethickness{1pt}
\epsfysize=38mm
\put(30,1){{\epsffile{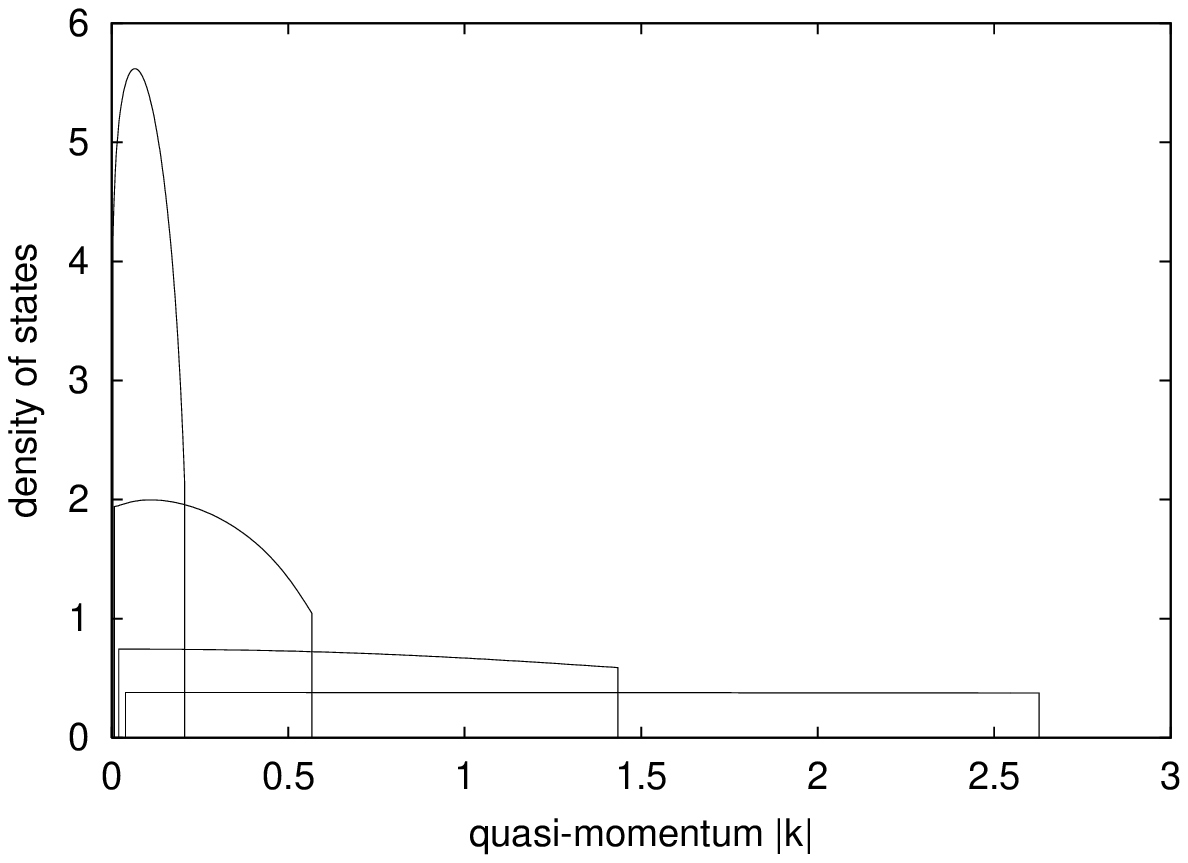}}}
\epsfysize=38mm
\put(88,1){{\epsffile{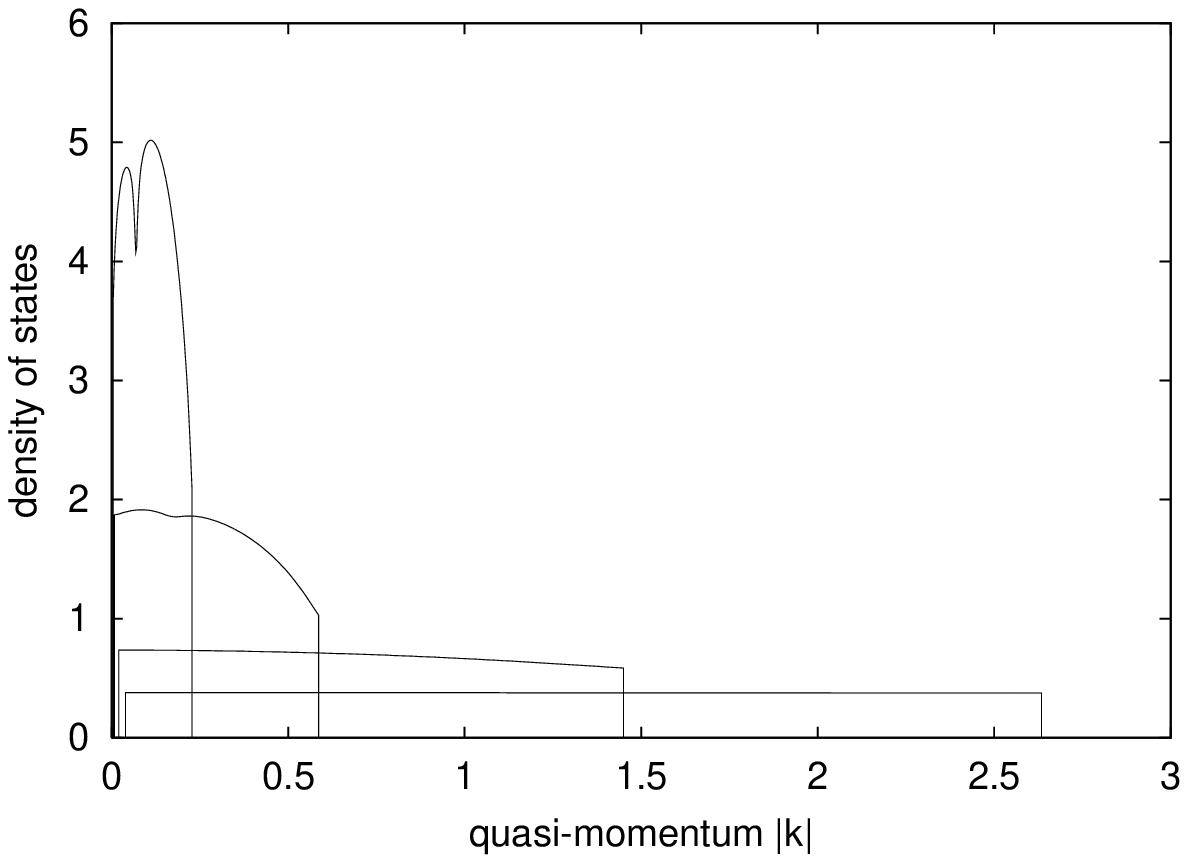}}}
\end{picture}
\vspace{0mm}
\caption{The density of state in $|k|$-space
for the ground state (left). The distribution
changes from a histogram to a narrow peak
gradually for the coupling from
strong to weak. 
The density of state in $|k|$-space in the
presence of one isospin rapidity at $J_1=30$ where a rift evident appeared (right). 
The figure is plotted for $N=L=100$ and $c=10,1, 0.1, 0.01$.}
\label{fig:ground}
\end{figure}

Similarly, from the first equation of eqs. (\ref{eq:secular-weak}) we have
\begin{equation}
2(I_{j+1}-I_j)-\frac{2L}{\pi}(k_{j+1}-k_j)-2=
-\sum_{\mu=1}^M[{\rm sgn}(k_{j+1}
-\lambda_\mu)-{\rm sgn}(k_{j}-\lambda_\mu)]
\label{GNN}
\end{equation}
For $I_{j+1}-I_j=n$, there will be $k_{j+1}-k_j=(n+m-1)\pi/L$
if there are $m$  $\lambda_\gamma$
satisfying $k_j<\lambda_\gamma<k_{j+1}$. Therefore, an isospin
rapidity of value $\lambda_\mu$ always repels the quasi-momentum
away from that value.
The ground state for $c\rightarrow 0$ does not allow any possible existence
of $\lambda_\mu$ between $k_1$ and $k_N$.
Thus from eq. (\ref{GMM1}) and eq. (\ref{GMM})
we can conclude that the total number of isospin down should be zero.
The quantum numbers $\{I^0_j\}$ of the ground state are given by
the successive positive numbers $I^0_1=1, I_{j+1}^0-I_j^0=1$.
In the thermodynamic limit, as
the difference between two adjacent $k$'s is small, the contribution
of any existing $\lambda$ to the density of $k$-rapidity at points
$k=\lambda$ will brings about a rift (see fig.\ref{fig:ground} right).
Therefore, an existing $\lambda_\mu$ will  suppress the density of
state in $k$-space at the point $k=\lambda_\mu$.
The more isospin rapidities there are, the higher the energy will be.
In contrast to the spin-$1/2$ Fermi system whose ground state is SU(2) singlet,
the ground state of trapped 2-component bosons
is an isospin ``ferromagnetic" state which is described by
the quantum number,
\begin{equation}
\{I_j\}:=\{1,\cdots,N\}, \; \{J_\gamma\}=empty
\label{GN}
\end{equation}
Since the existence of open boundary conditions makes
the backwards scattering $k\rightarrow -k$  possible, there
exists an additional term $\tan^{-1}(2k/c)$ which contributes
a negative value to $\rho(k)$, especially
$\rho(0)$ is obviously suppressed.
This phenomenon is different from that for the system with
periodic conditions. We interpret
it being due to the confinement by the infinite-depth
well, for example, the ground state for a single particle is
not $k=0$ but $k=\pi/L$ in open boundary conditions.
The density of states
of the ground state for various coupling is plotted in fig.\ref{fig:ground} 
the left.

\begin{figure}
\setlength\epsfxsize{55mm}
\begin{center}
\epsfbox{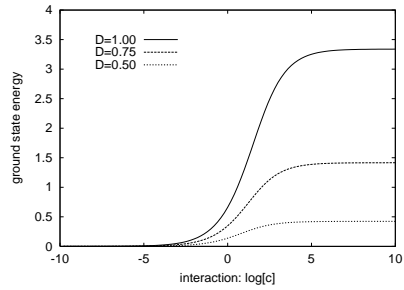}
\end{center}
\caption{The ground state energy versus the coupling
for different densities $D=1.0, 0.75, 0.5$.}
\label{fig:grd-density}
\end{figure}

By introducing
\begin{eqnarray}
L\rho\Bigr(\frac{k_j+k_{j+1}}{2}\Bigr)=&&\frac{1}{k_{j+1}-k_j} \nonumber \\
L\sigma\Bigr(\frac{\lambda_\gamma+\lambda_{\gamma+1}}{2}\Bigr)=&&
\frac{1}{\lambda_{\gamma+1}-\lambda_\gamma}
\label{RHOKD}
\end{eqnarray}
the density corresponding to the quantum numbers
(\ref{GN}) satisfies the integral equation
\begin{equation}
\rho_0(k)=\frac{1}{\pi}-\frac{1}{L}K_{1/2}(k)
 +\int_{0}^{Q_0}K_1(k|k')\rho_0(k')dk'
\label{eq:grddensity}
\end{equation}
where
$$
K_n(x)=\frac{1}{\pi}\frac{nc}{n^2c^2 + x^2}
$$
and
$$
K_n(x|y)=K_n (x-y)+K_n (x+y)
$$
$\rho_0, Q_0$ are respectively the density and integration limit
for the ground state.
The $K_{1/2}(k)$ term in (\ref{eq:grddensity}) comes from the fact that 
$\tan^{-1}(\frac{k_j-k_l}{c})\rightarrow 0$ but
$\tan^{-1}(\frac{k_j+k_l}{c})\rightarrow \tan^{-1}(\frac{k_j}{c/2})$
when taking account of  the term of $l=j$ in 
the process of thermodynamic limit.
The particle number per length is determined by
\begin{equation}
D=\frac{N}{L}=\int^{Q_0}_{0}\rho_0(k)dk
\label{eq:N1}
\end{equation}
eqs. (\ref{eq:grddensity}) and eq. (\ref{eq:N1}) determine $Q_0$ and 
$\rho_0(k)$. Then
the energy can be calculated by
\begin{equation}
\frac{E_0}{L}=\int_0^{Q_0}k^2\rho(k)dk
\end{equation}
which is explicitly  
$\displaystyle\frac{1}{3}\pi^2D^3(1-\frac{4}{c}D)$  
in the strong coupling limit $c\gg 1$.
In the general case one needs to solve the equations
numerically. We show the ground state energy for various densities
$D=1.0, 0.75, 0.5$ in fig.\ref{fig:grd-density}.

\section{Low-lying excited states}\label{sec:low}

\begin{figure}
\epsfclipoff
\fboxsep=88mm
\setlength{\unitlength}{1mm}
\begin{picture}(96,42)(0,0)
\linethickness{1pt}
\epsfysize=38mm
\put(30,1){{\epsffile{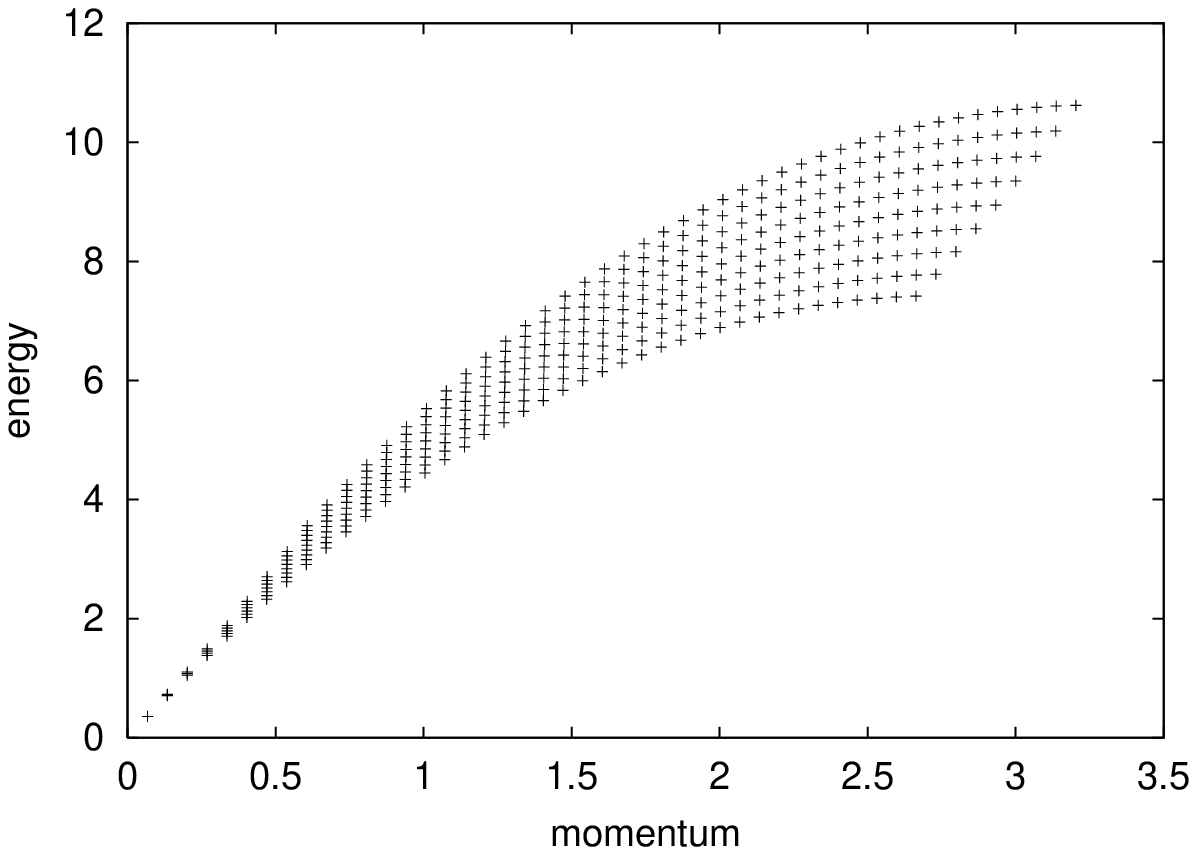}}}
\epsfysize=38mm
\put(88,1){{\epsffile{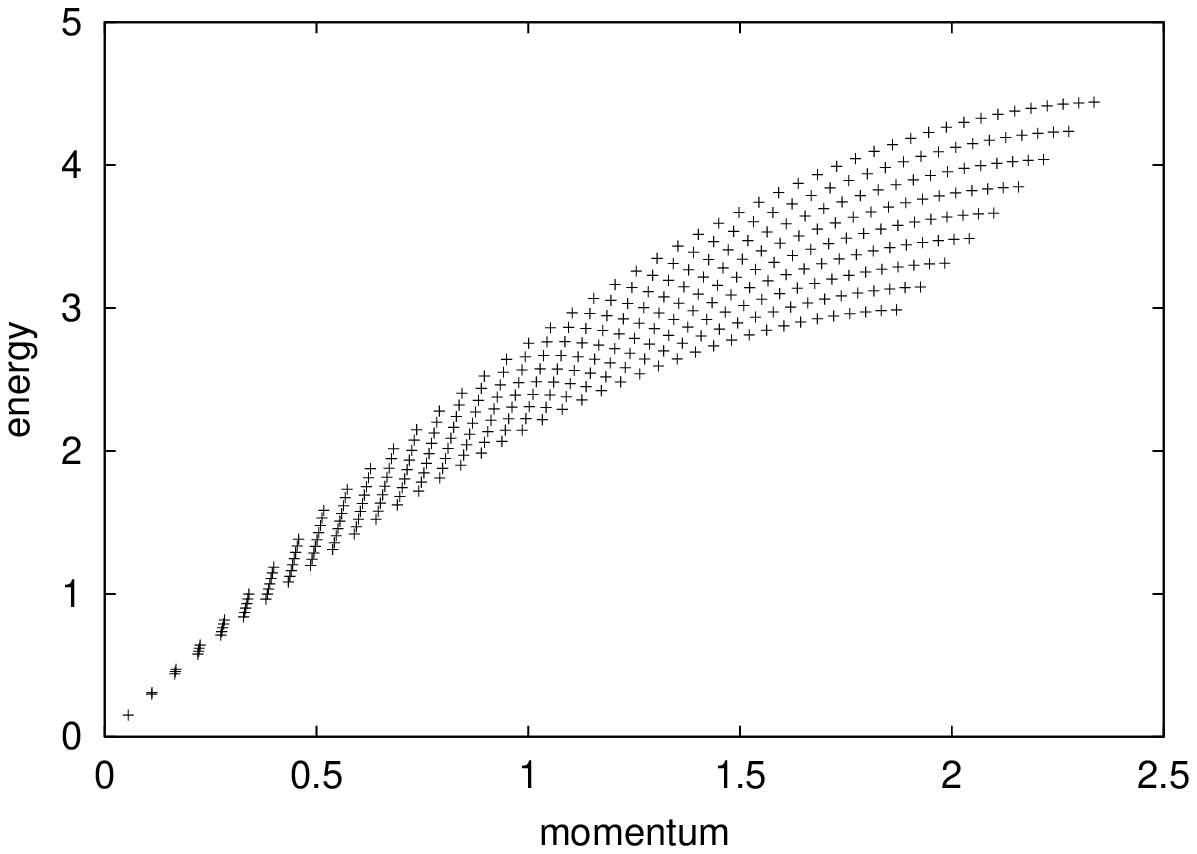}}}
\end{picture}
\vspace{0mm}
\caption{The holon-antiholon excitation spectrum
calculated for $N=L=40$ and $c=10$(left) and $c=1$(right).}
\label{fig:hole-ant}
\end{figure}

\begin{figure}
\epsfclipoff
\fboxsep=88mm
\setlength{\unitlength}{1mm}
\begin{picture}(76,42)(0,0)
\linethickness{1pt}
\epsfysize=38mm
\put(30,1){{\epsffile{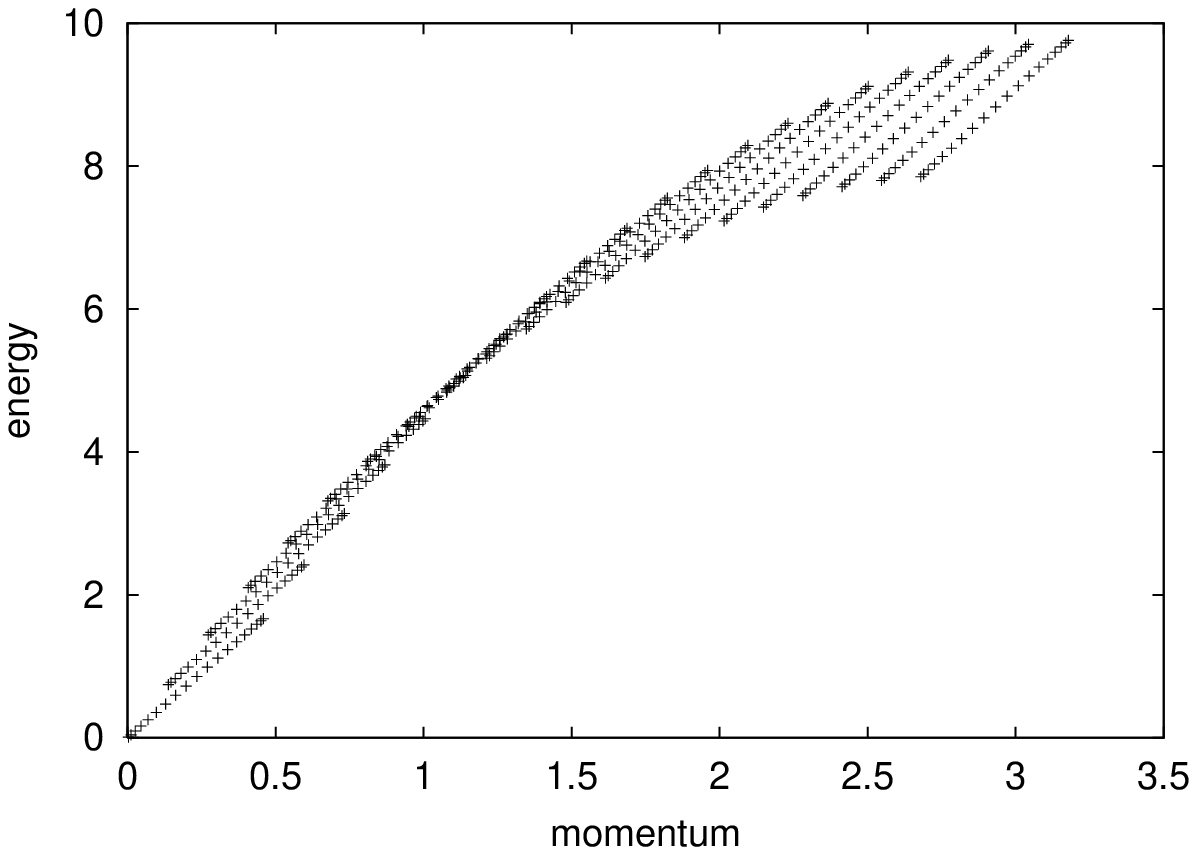}}}
\epsfysize=38mm
\put(88,1){{\epsffile{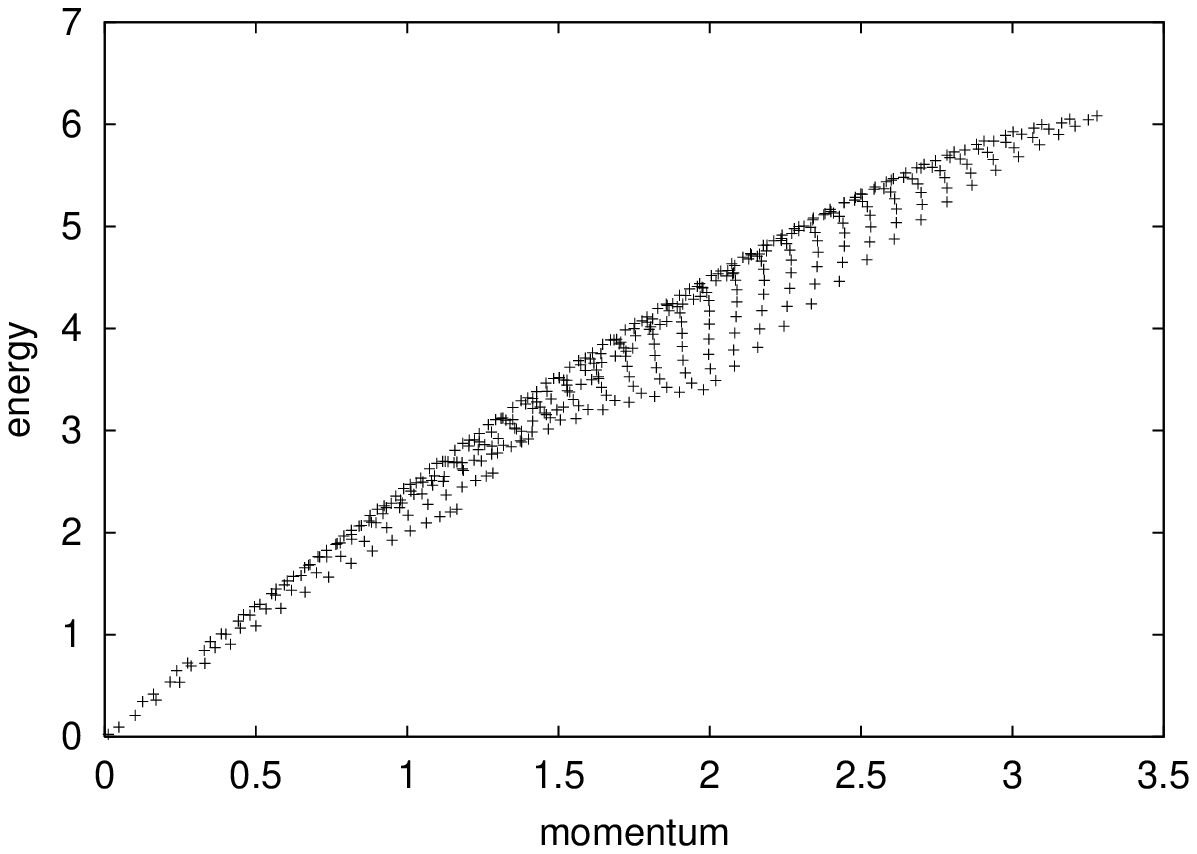}}}
\end{picture}
\vspace{0mm}
\caption{The holon-isospinon excitation spectrum
calculated for $N=L=20$, $c=10$ (left) and $c=1$(right).}
\label{fig:hole-iso}
\end{figure}

\begin{figure}
\setlength\epsfxsize{55mm}
\begin{center}
\epsfbox{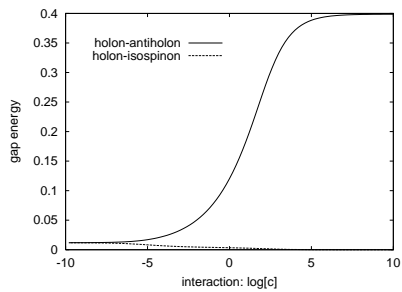}
\end{center}
\caption{The finite-size energy gap versus interaction for $N=L=50$,
obviously, $c=0$ corresponds to 0.02=1/L}
\label{fig:gap-log}
\end{figure}

The excited states are obtained by the variation of
the configuration $\{I_j, J_\gamma\}$ from that
of the ground state. The simplest case is to
remove one of $Is$ from the configuration of the ground state
and add a new $I_n$ outside the original sequence. We call
this holon-antiholon excitation which is described by
$$
\{I_j\}=\{1,\cdots,n-1, n+1,\ldots,N, I_n \}
$$
with $|I_n|>N$ and $M=0$.
To investigate the excited states, we shall consider
systems of finite size first, then take
thermodynamics limits.
Although the total momentum $\sum_j k_j$ is no more a constant
in open boundary conditions, $\sum_j |k_j|$ is still a constant
for a given energy eigenstate.
We plotted the numerical results for energy-``momentum'' spectrum
in fig.\ref{fig:hole-ant},
and \ref{fig:hole-iso} for $N=40,L=40$ where the $x$-axis represents the
quantitive change $\sum_j (|k_j|-|k^0_j|)$. We notice
that the overall structure of the spectrum does not change substantially
between weak ($c=1$) and strong ($c=10$) coupling regimes.
For a finite-size system, the gap of holon-antiholon
excitation opens, and it's dependence on the inter-particle interaction
is shown by the upper line of fig.\ref{fig:gap-log}.

In the thermodynamic limit, it is plausible to calculate
the excitation energy by making
$\rho(k)=\rho_0(k)+\rho_1(k)/L$ where $\rho_0(k)$
is the density of the ground state. By creating a
hole inside the quasi Fermi sea $\bar{k}\in [0, Q_0]$ and an
additional $k_p>Q_0$ outside it, we have 
\begin{equation}
\rho_1(k)+\delta(k-\bar k)=\int_{0}^{Q_0}dk^\prime
\rho_1(k^\prime)K_1(k|k^\prime)+K_1(k|k_p)
\end{equation}
The excitation energy consists of two terms
$\triangle E=\int k^2 \rho(k)dk-\int k^2 \rho_0(k)dk
=\varepsilon_h({\bar k})+\varepsilon_a(k_p)$.
The holon energy $\varepsilon_h$ and antiholon energy
$\varepsilon_a(k_p)=-\varepsilon_h(k_p)$ are given by
\begin{eqnarray}
\varepsilon_h({\bar k})=-{\bar k}^2
  +\int_0^{Q_0}k^2\rho_1^h(k,{\bar k})dk, \nonumber \\
\rho_1^h(k,{\bar k})=-K_1(k|{\bar k})
   +\int_0^{Q_0}K_1(k|k^\prime)\rho_1^h(k^\prime|{\bar k})
\label{HAE}
\end{eqnarray}

\begin{figure}
\setlength\epsfxsize{55mm}
\begin{center}
\epsfbox{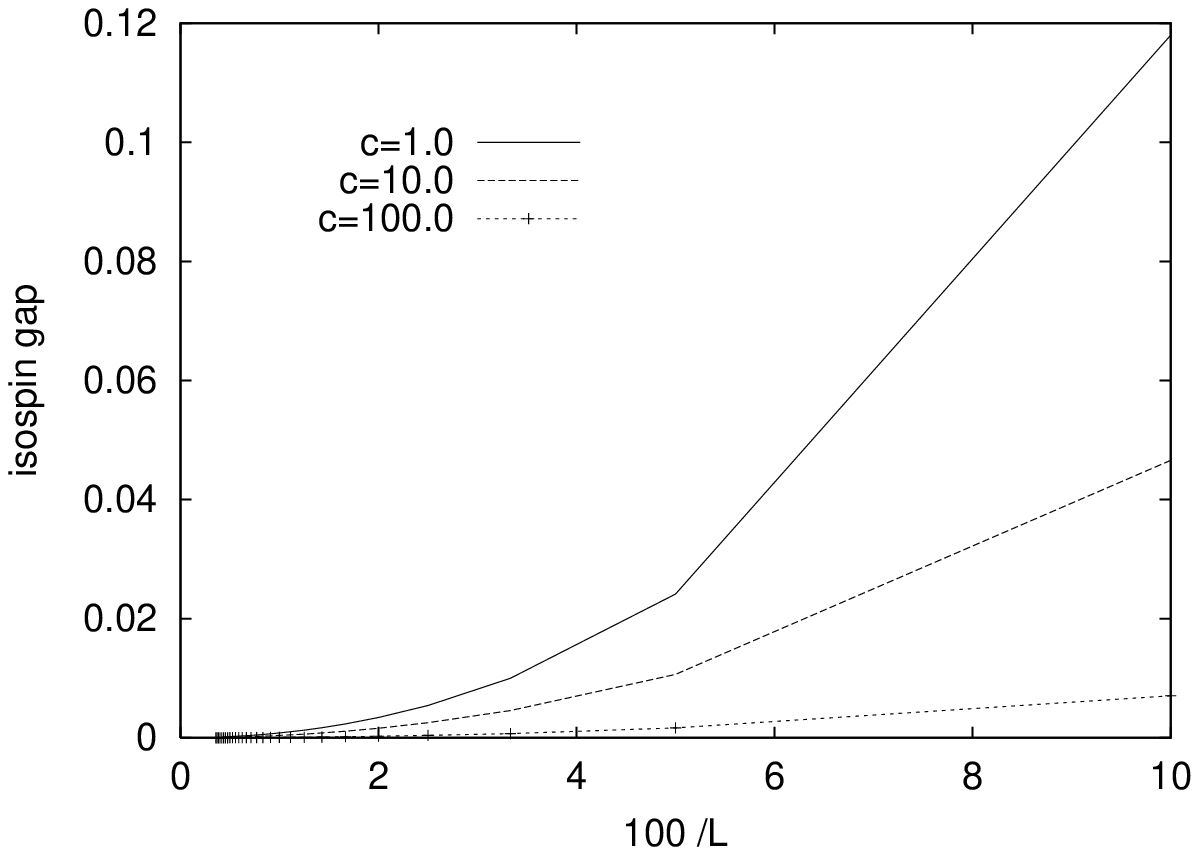}
\end{center}
\caption{Finite-size effects: Isospin gap as a function of 100/L
for $D=N/L=1$, plotted for various inter-particle interaction}
\label{fig:gap-iso}
\end{figure}

Another interesting excitation is to
flip one isospin down, i.e.,
$M=1$.
As discussed above, a single $\lambda$
can suppress the density of state
at $k=\lambda$.
We plot the density of the lowest energy state for
various couplings in fig.\ref{fig:ground} the right
with $N=100, L=100$ and $J_1=30$.
The holon-isospinon excitation is characterized
by the Young tableau [$N-1, 1$],
accordingly, $I$'s take $N$ distinct
integers and $I_1<J_1<I_N$, namely
$$
I_1=-N/2+\delta_{1,j_1}\;\;\;(1\leq j_1\leq N+1)
$$
$$
I_j=I_{j-1}+1+\delta_{j,j_1}\;\;(j=2,\dots,N)
$$
where $\delta_{\alpha,\beta}$ is the usual Kronecker
symbol. The excitation spectra are plotted in
fig.\ref{fig:hole-iso}, for
$L=20, N=20, c=10$ and $c=1$ respectively.
The lowest energy state of one isospin
down is to make the absolute value of
$J_1$ as large as possible in order to
avoid the enhancement of energy caused by the suppression of
the $k$-distribution.
In the thermodynamic limit, it can be shown that
this mode is gapless.
For a finite particle system, however, it still
has a gap of order $1/L$ which decreases more quickly than
that for periodic conditions for particle number
ranging from small to large.
Figure \ref{fig:gap-log} shows it's dependence of
the gap on the interaction. We presented in
fig.\ref{fig:gap-iso} the behavior of the
finite-size spin gap as a function of $1/L$ at $c=1.0, 10.0, 100.0$.

In the thermodynamic limit, we should take into account of
the hole in the $k$-sector and the $\rho_1(k)$ should be solved from
\begin{equation}
\rho_1(k)+\delta(k-{\bar k})=\int K_1
(k|k^\prime)\rho_1(k^\prime)dk^\prime
-K_{1/2}(k|\lambda)
\end{equation}
One can find that the energy of holon-isospinon excitation
consists of two terms
$\triangle E=\varepsilon_h({\bar k})+\varepsilon_{iso}(\lambda)$.
The $\varepsilon_h$ is determined by
eqs. (\ref{HAE}) and the $\varepsilon_{iso}$
by $\varepsilon_{iso}(\lambda)=\int k^2\rho_1^{iso}(k,\lambda)dk$
with
\begin{equation}
\rho_1^{iso}(k,\lambda)=-K_{1/2}(k|\lambda)+\int_0^{Q_0}
 K_1(k|k')\rho_1^{iso}(k',\lambda)dk'
\end{equation}

\section{Thermodynamics at finite temperature}\label{sec:thermo}

For the ground state (i.e. at zero temperature),
the charge rapidity $k$'s are real roots of the Bethe-ansatz
equations (\ref{eq:BAE}). For the excited state, however, the isospin
rapidity can be complex roots \cite{Takahashi} which always
form a ``bound state" with several other $\lambda$'s.
This arises from the consistency of both sides of the Bethe-ansatz
equations \cite{Woynarovich}. The {\it n-string} of $\lambda$ rapidity is
defined as
\begin{equation}
\Lambda_a^{nj}=\lambda_a^n+(n+1-2j)ic/2+O(\exp(-\delta N))
\label{stringdefine}
\end{equation}
where $j=1, 2\cdots n$. The number of total down isospin
is determined by
\begin{equation}
M=\sum_{n=1}^\infty nM_n
\end{equation}
where $M_n$ denotes the number of n-strings. The eqs. (\ref{eq:secular})
become
\begin{eqnarray}
\pi I_j&=&k_j L
          +\sum_{l\neq j}^N\Bigl[\tan^{-1}\Bigr(\frac{k_j-k_l}{c}\Bigr)
           +\tan^{-1}\Bigr(\frac{k_j+k_l}{c}\Bigr)\Bigr]
            \nonumber \\
   &\,& -\sum_{an}\Bigl[\tan^{-1}\Bigr(\frac{k_j-\lambda_a^n}{nc/2}\Bigr)
      +\tan^{-1}\Bigr(\frac{k_j+\lambda_a^n}{nc/2}\Bigr)\Bigr]
        \nonumber \\
\pi J_a^n &=&\tan^{-1}\Bigr(\frac{\lambda_a^n}{nc/2}\Bigr)
   +\sum_{l=1}^N\Bigr[\tan^{-1}\Bigr(\frac{\lambda_a^n-k_l}{nc/2}\Bigr)
    +\tan^{-1}\Bigr(\frac{\lambda_a^n+k_l}{nc/2}\Bigr)\Bigr]
       \nonumber \\
  &\,& -\sum_{mbt}A_{mnt}\Bigr[\tan^{-1}
    \Bigr(\frac{\lambda_a^n-\lambda_b^m}{tc/2}\Bigr)
    +\tan^{-1}\Bigr(\frac{\lambda_a^n+\lambda_b^m}{tc/2}\Bigr)\Bigr]
\label{seqular}
\end{eqnarray}
where
\[
A_{nmt}=\left\{\begin{array}{ll}
1, & {\rm for}\; t=m+n, |m-n|(\neq 0)\\
2, & {\rm for}\; t=n+m-2,\cdots,|n-m|+2\\
0,  & {\rm  otherwise}
\end{array}\right.
\]
and the quantum numbers $\{I_j, J_a^n\}$ label
the state beyond the ground state.
The densities corresponding to charge rapidity and isospin rapidity
on the real axis for which
the ``omitted $k, \lambda$ values'' must be taken into account.
By introducing the density
of holes for charge $\rho^h(k)$ and isospin n-string
$\sigma_n^h(\lambda)$, we have
\begin{eqnarray}
\rho+\rho^h=&&\frac{1}{\pi}-\frac{1}{L}K_{1/2}(k)
+\int K_1(k|k^\prime)\rho(k^\prime)dk'
\nonumber \\
&&-\sum_n\int K_{n/2}(k|\lambda)\sigma_n(\lambda)d\lambda
\nonumber \\
\sigma_n+\sigma_n^h=&&\frac{1}{L}
K_{n/2}(\lambda)+\int K_{n/2}(\lambda|k)\rho(k)dk
\nonumber \\
&&-\sum_{mt}A_{mnt}\int K_{t/2}(\lambda|\lambda')\sigma_m(\lambda')d\lambda'
\label{density}
\end{eqnarray}
where the integration limits are $[0,\infty]$.
In terms of the distribution functions
of charge and isospin rapidities,
the energy per length has the form $E_k/L=\int k^2\rho(k)dk$,
the total number of down isospins is
$M/L=\sum_n n\int\sigma_n(\lambda)d\lambda$ and
the particle density of the system is $D=N/L=\int \rho(k) dk$.

If we consider the energy arising from the external field $\Omega$ which
is the Rabi field in two-component BEC experiments,
the internal energy of the system is
\begin{equation}
E/L=\int(k^2-\Omega)\rho(k)dk+\sum_n 2n\Omega\int\sigma_n d\lambda
\end{equation}
For a given $\rho(k)$, $\rho^h(k)$, $\sigma_n(\lambda)$
and $\sigma_n^h(\lambda)$, the entropy
is of the form \cite{Yang}
\begin{eqnarray}
S/L=&&\int [(\rho+\rho^h)\ln(\rho+\rho^h)
 -\rho\ln\rho-\rho^h\ln\rho^h]dk
    \nonumber\\
&&+\sum_n\int[(\sigma_n+\sigma_n^h)\ln(\sigma_n+\sigma_n^h)
 -\sigma_n\ln\sigma_n
  -\sigma_n^h\ln\sigma_n^h]d\lambda
\end{eqnarray}
where the Boltzmann constant is set to unity.

At finite temperature, we should minimize
the free energy $F=E-TS-\mu N$ where $\mu$
is the chemical potential and $S$ is the
entropy of the system. Making use of the
relations derived from eqs. (\ref{density})
\begin{eqnarray}
\delta\rho^h=&&-\delta\rho+\int
K_1(k|k')\delta\rho dk'
   -\sum_n\int K_{n/2}(k|\lambda)
   \delta\sigma_n d\lambda
\nonumber \\
\delta\sigma_n^h=&&-\delta\sigma_n
+\int K_{n/2}(\lambda|k)\delta\rho dk
  -\sum_{mt}A_{mnt}\int K_{t/2}(\lambda|\lambda^\prime)
   \delta\sigma_m(\lambda^\prime)d\lambda^\prime
\end{eqnarray}
we obtain the following conditions from the
minimum condition $\delta F=0$,
\begin{eqnarray}
\epsilon(k)&&=k^2-\Omega-\mu-T\int K_1(k|k')
 \ln(1+e^{-\epsilon(k^\prime)/T})dk'
   \nonumber \\
&&-T\sum_n\int K_{n/2}(k|\lambda)
  \ln[1+e^{-\zeta_n(\lambda)/T}]d\lambda
   \nonumber \\
\zeta_n(\lambda)&&=2n\Omega+T\int K_{n/2}(k|
  \lambda)\ln[1+e^{-\epsilon(k)/T}]dk
   \nonumber \\
&&+T\sum_{mt}A_{mnt}\int K_{t/2}(\lambda|\lambda^\prime)
   \ln[1+e^{-\zeta_m(\lambda^\prime)/T}]d\lambda^\prime
\label{thermalequations}
\end{eqnarray}
where we have written
\begin{eqnarray}
\frac{\rho^h(k)}{\rho(k)}=&&e^{\epsilon(k)/T}
\nonumber \\
\frac{\sigma_n^h(\lambda)}{\sigma_n(\lambda)}
=&&e^{\zeta_n(\lambda)/T}.
\end{eqnarray}

When $T\rightarrow 0$, eqs. (\ref{thermalequations}) become
\begin{eqnarray}
{\cal E}_0(k)=&&k^2-\Omega-\mu+\int K_1(k|k')
  {\cal E}_0(k')dk'
\label{DRE}
\end{eqnarray}
where the ``ferromagnetic'' ground state is under consideration.
Clearly, the integral equation gives the solution
for the dressed energy \cite{Frahm} from which the
ground-state energy can be given in terms of
${\cal E}_0$
\begin{equation}
E_0/L=\frac{1}{\pi}\int_{0}^{Q_0}{\cal E}_0(k)dk
\end{equation}
The solution of eq. (\ref{DRE}) defines the energy bands.
And the Fermi surface is determined by
\begin{equation}
{\cal E}_0(k_F)=0
\end{equation}
since the ground-state configuration corresponds
to the case that all states of ${\cal E}_0(k)<0$ are fully filled.
The bare energy ${\cal E}^{(0)}_0$ is the zero
order term of eq. (\ref{DRE})
\begin{equation}
{\cal E}^{(0)}_0=k^2-\Omega-\mu
\end{equation}
which is valid in the strong coupling limit

Equations (\ref{thermalequations}) can be solved  by iteration.
The coupled equations (\ref{density})
of density distribution of charge and isospin are then
a Fredholm type
\begin{eqnarray}
\rho(1+e^{\epsilon/T})=&&\frac{1}{\pi}-\frac{1}{L}K_{1/2}(k)
  +\int K_1(k|k')\rho(k')dk'
   \nonumber \\
&&-\sum_n\int K_{n/2}(k|\lambda)
\sigma_n(\lambda)d\lambda, \nonumber \\
\sigma_n(1+e^{\zeta_n/T})=&&\frac{1}{L}K_{n/2}(\lambda)
+\int K_{n/2}(\lambda|k)\rho(k)dk
\nonumber \\
&&-\sum_{mt}A_{mnt}\int K_{t/2}(\lambda|\lambda^\prime)
\sigma_m(\lambda^\prime)d\lambda^\prime
\end{eqnarray}
Finally, we obtain the Helmholtz free energy $F=E-TS$
\begin{eqnarray}
F=&&\mu N-T\int\ln[1+e^{-\epsilon/T}]
  \Bigl[\frac{L}{\pi}-K_{1/2}(k)\Bigr]dk  \nonumber \\
  &&-T\sum_n\int\ln[1+e^{-\zeta_n/T}]K_{n/2}(\lambda)d\lambda
\label{GFE}
\end{eqnarray}
and the pressure
\begin{equation}
P=-\frac{\partial F}{\partial L}=\frac{T}{\pi}\int\ln[1+e^{-\epsilon/T}]dk,
\end{equation}
which is formally the same as Yang and Yang's expression but the
equations determining $\epsilon$ and $\zeta$ are different.
Consequently, the partition function is given
by $Z=e^{-F/T}$. The thermodynamic
functions, such as partition function $Z$,
free energy $F$, are of importance
for a thermal system. Given either
of them, one is able to calculate all
thermodynamic properties for the
system in principle. However, the
eqs. (\ref{thermalequations})
are so complicated that it is impossible
to obtain the explicit form of
$\epsilon$ and $\zeta$ for general
case. Moreover, we can also obtain
some plausible results for some special
cases in the next section.

\section{Special cases}\label{sec:spec}

In general, the free energy of our model should
be calculated using formula (\ref{GFE}),
where $\epsilon(k)$ and $\zeta_n(\lambda)$ are
determined from eqs. (\ref{thermalequations}).
Eqs. (\ref{thermalequations}) can be solved
by numerical method via iteration.
In this section, we will only discuss the
case of strong coupling at low temperature.

When $c\rightarrow \infty$, $K_n(k|k^\prime)=0$
and $K_n(k)=0$, then we have
\begin{equation}
\epsilon=k^2-\Omega-\mu
\end{equation}
and the free energy becomes
\begin{equation}
F/L=\mu D-\frac{T}{\pi}\int\ln[1+e^{-\epsilon/T}]dk
\end{equation}
which can be solved by integration by part \cite{GuLL},
\begin{equation}
F/L=\mu D-\frac{2}{\pi}
  \Bigr(\frac{1}{3}\mu^{3/2}+\frac{T^2\pi^2}{24\mu^{1/2}}\Bigr)
\end{equation}
where the external field is set to zero.

We can not derive the specific heat
directly from the afore-obtained free energy
because the chemical potential
is a function of temperature.
From eqs. (\ref{density}), the density
of charge rapidity has the form
\begin{equation}
\rho=\frac{1}{\pi}
\frac{1}{1+e^{(k^2-\Omega-\mu)/T}}
\end{equation}
Clearly, at zero temperature, the quasi-Fermi
surface is just the square  root of the
chemical potential, so we have
$\mu_0=\pi^2 D^2$
which denotes the chemical potential at
zero temperature. At low temperature,
however, it is determined by
\begin{equation}
D=\frac{1}{\pi}
   \int_{0}^\infty\frac{1}{1+e^{(k^2-\mu)/T}}dk
\end{equation}
we have
\begin{equation}
\mu=\pi^2 D^2\Bigr[1+\frac{\pi^2 T^2}{24 \mu^2}\Bigr]^{-2}
\end{equation}
The second term in the bracket is small at low temperature, so we
can replace $\mu$ by $\mu_0$. The equation becomes
\begin{equation}
\mu=\mu_0\Bigr[1-\frac{\pi^2 T^2}{24\mu_0^2}\Bigr]^{-2}
\end{equation}
then the free energy becomes
\begin{equation}
F/L=\mu_0 D\Bigr[1+\frac{\pi^2T^2}{12\mu_0^2}\Bigr]
  -\frac{2}{3\pi}
  \mu_0^{3/2}\Bigr[1+\frac{\pi^2T^2}{4\mu_0^2}\Bigr]
\end{equation}

Since by thermodynamics $S=-\partial F/\partial T$ and
$C_v=T\partial S/\partial T$, we find the specific heat
at low temperature is Fermi-liquid like
\begin{equation}
S=C_v=\frac{T}{6D}
\end{equation}
It is the same as the result of the one-component case,
since for the strong coupling limit the isospin and charge are decoupled,
the contribution of isospinon to the free energy vanishes.
In fig.\ref{fig:gap-iso}, the finite-size energy gap of holon-isospinon
in strong coupling limit tends to zero.

\section{Conclusions}

In this paper, we have solved a system of
one dimensional trapped SU(2) bosons with
$\delta$-function interaction by means of the
Bethe-ansatz method. On the basis of Bethe-ansatz
equations we first discussed the ground state of
the system and found that the ground state
is an isospin ``ferromagnetic" state which
differs greatly from the spin-1/2 fermion systems.
We studied the low excitation states by
both numerical and analytic methods.
It was shown that there are three elementary excitation modes, and
the holon-antiholon and holon-isospinon excitations are gapless
for large systems.
For finite system, we not only plotted some excitation spectra
but also plotted the dependence of finite-size energy gap on the
inter-particle interaction.
The thermodynamics of the
system were studied by using the thermodynamic Bethe ansatz \cite{Yang}.
For strong coupling we found that the system exhibits
the Fermi-liquid behavior, i.e. the specific heat is a
linear function of $T$ at low temperature.

\section*{Acknowledgment}

This work is supported by Trans-Century Training Program Foundation
for the Talents and EYF98 of China Ministry of Education.
YQL is also supported by AvH-Stiftung.
SJG thanks D.Yang and YQL thanks K.Marzlin for helpful discussions.
We would like to thank the referees for their helpful suggestions.


\section*{References}

\begin{thebibliography}{99}

\bibitem{Bagnato}
Bagnato V and Kleppner D 1991 {\it Phys. Rev. } {\bf A 44} 7439

\bibitem{Ketterle}
Ketterle W and Druten N J 1996 {\it Phys. Rev.} {\bf A 54} 656

\bibitem{Druten}
Druten N J and Ketterle W 1997 {\it Phys. Rev. Lett.} {\bf 79} 549

\bibitem{LiebL}
Lieb E H and Liniger W 1963 {\it Phys. Rev.} {\bf 130} 1605
Lieb E H 1963 {\it Phys. Rev.} {\bf 130}, 1616

\bibitem{Haldane}
Haldane F D M 1981 {\it Phys. Rev. Lett.} {\bf 47} 1840

\bibitem{Monien}
Monien H, Linn M and Elstner N 1998 {\it Phys. Rev.} {\bf A 58} R3395

\bibitem{Williams1}
Williams J E and Holland M J 1999 {\it Nature} {\bf 401} 568

\bibitem{Williams2}
Williams J E, Holland M J, Wieman C E and Cornell E A 1999
{\it Phys. Rev. Lett.} {\bf 83} 3358

\bibitem{Ho}
Ho T L 1998 {\it Phys. Rev. Lett.} {\bf 81} 742

\bibitem{LiGYE}
Li Y Q, Gu S J, Ying Z J and Eckern U submitted to PRL

\bibitem{Takahashi}
Takahashi M 1971 {\it Prog. Theor. Phys.} {\bf 46} 1388

\bibitem{Woynarovich}
Woynarovich F 1982 {\it J. Phys. C: Solid State Phys.} {\bf 15} 85\\
Woynarovich F 1982 {\it J. Phys. C: Solid State Phys.} {\bf 15} 97

\bibitem{Yang}
Yang C N and Yang C P 1969 {\it J. Math. Phys.} {\bf 10} 1115

\bibitem{Li}
Li Y Q 1995 {\it Phys. Rev.} A {\bf 52}, 65 \\
Li Y Q and Gruber C 1998 {\it Phys. Rev. Lett} {\bf 80} 1034

\bibitem{LDFaddeev84}
Faddev L D, in {\it Recent Advances in Field Theory and Statistical
Mechanics}, ed. Zuber J and Stora (Elsevier, Amsterdam 1984) p. 569.

\bibitem{Gaudin}
Gaudin M 1971 {\it Phys. Rev. A} {\bf 4} 386 (1971)

\bibitem{Frahm}
Frahm H and Korepin V E 1990 {\it Phys. Rev.} {\bf B42} 10553

\bibitem{GuLL}
Gu S J, Li Y Q and Lin H Q 2000 {\it J. Phys. A: Math. Gen.} {\bf 33} 6779

\end{thebibliography}
\end{document}